\documentclass[manuscript,screen,nonacm]{acmart}

\usepackage{url,hyperref,lineno,microtype,subcaption}
\usepackage{braket}

\usepackage{listings}
\usepackage{xcolor}
\usepackage{amsmath, xparse}

\definecolor{codegreen}{rgb}{0,0.6,0}
\definecolor{codegray}{rgb}{0.5,0.5,0.5}
\definecolor{codepurple}{rgb}{0.58,0,0.82}
\definecolor{backcolour}{rgb}{1,1,1}

\lstdefinestyle{mystyle}{
    backgroundcolor=\color{backcolour},   
    commentstyle=\color{codegreen},
    keywordstyle=\color{magenta},
    numberstyle=\tiny\color{codegray},
    stringstyle=\color{codepurple},
    basicstyle=\ttfamily\footnotesize,
    breakatwhitespace=false,         
    breaklines=true,                 
    captionpos=b,                    
    keepspaces=true,                 
    numbers=left,                    
    numbersep=5pt,                  
    showspaces=false,                
    showstringspaces=false,
    showtabs=false,                  
    tabsize=2
}

\AtBeginDocument{%
  \providecommand\BibTeX{{%
    \normalfont B\kern-0.5em{\scshape i\kern-0.25em b}\kern-0.8em\TeX}}}

\begin{document}

\title{On Physics-Informed Neural Networks for Quantum Computers}

\author{Stefano Markidis}
\email{markidis@kth.se}
\affiliation{%
  \institution{KTH Royal Institute of Technology}
  \city{Stockholm}
  \country{Sweden}
}

\renewcommand{\shortauthors}{Markidis}

\begin{abstract}
Physics-Informed Neural Networks (PINN) emerged as a powerful tool for solving scientific computing problems, ranging from the solution of Partial Differential Equations to data assimilation tasks. One of the advantages of using PINN is to leverage the usage of Machine Learning computational frameworks relying on the combined usage of CPUs and co-processors, such as accelerators, to achieve maximum performance. This work investigates the design, implementation, and performance of PINNs, using the Quantum Processing Unit (QPU) co-processor. We design a simple Quantum PINN to solve the one-dimensional Poisson problem using a Continuous Variable (CV) quantum computing framework. We discuss the impact of different optimizers, PINN residual formulation, and quantum neural network depth on the quantum PINN accuracy. We show that the optimizer exploration of the training landscape in the case of quantum PINN is not as effective as in classical PINN, and basic Stochastic Gradient Descent (SGD) optimizers outperform adaptive and high-order optimizers. Finally, we highlight the difference in methods and algorithms between quantum and classical PINNs and outline future research challenges for quantum PINN development.
\end{abstract}


\maketitle

\section{Introduction}
\label{intro}
One of the most exciting and lively current research topics in scientific computing is integrating classical scientific methods with Machine Learning (ML) and neural network approaches~\citep{markidis2021old}. The usage of Physics-Informed Neural Networks (PINNs) is a major advancement in this promising research direction. PINNs have been applied in a wide range of traditional scientific computing applications, ranging from Computational Fluid Dynamics~(CFD)~\citep{cai2022physics} to solid mechanics~\citep{haghighat2021sciann}, and electromagnetics~\citep{chen2020physics}, to mention a few PINN usages. 

In essence, PINNs are neural networks that allow solving a Partial Differential Equation (PDE) of a specific domain area, such as Navier-Stokes equations for CFD or the Poisson equation in electrostatic problems. To achieve this, PINNs combine and connect two neural networks: a surrogate and a residual network. The first surrogate neural network takes as input the point we want to calculate the PDE at (this point is called \emph{collocation point}) and provides the approximated solution at that point. Using a training process, the surrogate neural network encodes the Partial Differential Equation (PDE) and associated boundary and initial conditions as weights of biases. The second residual network (not to be confused with the popular ResNet!) takes the input from the surrogate network. It provides the PDE \emph{residual}, which is the error of the approximated solution from the surrogate network. The residual calculation requires the solution of differential operators on a neural network. This calculation is performed using automatic differentiation~\citep{baydin2018automatic} that allows calculating differential operators at any given point without discretization and meshes (PINNs are gridless methods). The residual network has only a passive role during the training process. While it provides the surrogate network with the loss function, its weights and biases are not updated during the neural network training. A stochastic optimizer, such as the Stochastic Gradient Descent (SGD)~\citep{amari1993backpropagation} and Adam~\citep{kingma2014adam}, updates the surrogate network weights and biases using the residual obtained from the residual network. Typically, in PINNs, we use a sequence of Adam and Limited-memory Broyden–Fletcher–Goldfarb–Shanno (BFGS)~\citep{liu1989limited} optimizers: the L-BFGS optimizer is used to speed up the training or accelerate the convergence to the solution. PINNs do not require labeled datasets for the training, work in an unsupervised fashion, and necessitate only expressing the PDE in the residual network. For an in-depth description of the PINN, we refer the interested reader to the seminal work on PINNs by~\cite{raissi2019physics} and the following enlightening articles~\citep{lu2021deepxde, shin2020convergence, mishra2022estimates}.

From the computational point of view, one of the strategic advantages of using PINN methods for scientific computing is the possibility of exploiting high-performance frameworks, such as TensorFlow~\citep{abadi2016tensorflow} or PyTorch~\citep{paszke2019pytorch}, designed specifically for efficiently running ML workloads. Such frameworks rely heavily on Graphics Processor Units (GPUs) and accelerators for performance~\citep{chien2019tensorflow}. The use of GPUs for forward and back propagations led to large performance improvement and a renaissance of the research on neural networks, spurring the development of new deeper neural architectures. In addition to the performance, the domain scientist is not burdened with learning relatively low-level programming approaches, such as OpenCL or CUDA, but simply can pin computation to a device or rely on compiler technologies for accelerator automatic code generation~\citep{lattner2020mlir}. However, with Dennard’s scaling ending in 2005~\citep{horowitz2005scaling} and Moore’s law~\citep{moore1998cramming} possibly on its last days, many-core architectures (including GPUs) may not be enough to improve the performance scaling in a post-Moore era~\citep{theis2017end}. For this reason, researchers and companies are exploring alternative, disruptive computing directions, such as quantum computing, to further scale the performance beyond the limitation of silicon-based hardware. For instance, companies, such as Google and IBM, that invested heavily in the past in silicon-based accelerator technologies for AI workloads, now also investigate the development of quantum hardware and software for supporting ML workloads. Notable examples are the IBM-Q devices~\citep{chow2021ibm} and Qisikit~\citep{mckay2018qiskit} for IBM, the Sycamore quantum computer~\citep{arute2019quantum}, and Quantum TensorFlow~\citep{broughton2020tensorflow} for Google.

This work aims to investigate the potential of using an emerging co-processor, the Quantum Processing Unit (QPU), and associated software to deploy PINN on quantum computers. Quantum computing is an emerging technology and computational model for solving applications in various fields, ranging from cryptology~\citep{gidney2021factor} to database search~\citep{grover1996fast} to quantum chemistry simulations~\citep{o2016scalable}. Among these applications are traditional scientific computing and the solution of differential equations. These basic solvers are at the backbone of CFD, electromagnetics, and chemistry, among others. Algorithms and methodologies aimed at solving linear systems started with the work by~\cite{harrow2009quantum}, the development of the Harrow-Hassidim-Lloyd (HHL). They continued with linear solvers based on variational quantum solvers~\citep{bravo2019variational}, the seminal work on Differentiable Quantum Circuits (DQCs) for the solution of differential linear and non-linear equations~\citep{kyriienko2021solving, paine2021quantum,heim2021quantum, kyriienko2022protocols,kumar2022integral}, solvers with quantum kernel methods~\citep{paine2022quantum}, and linear systems based on quantum walks~\citep{chen2019hybrid}. On the road to fault-tolerant universal quantum computing systems, Noisy Intermediate-Scale Quantum (NISQ) systems~\citep{preskill2018quantum}, are currently major candidate systems for the design and development of near-term applications of quantum computing. Algorithms for NISQ systems are heterogeneous approaches as they combine code running on CPU (typically an optimizer or variational solver) and QPU (for a cost function evaluation). In this work, we focus on a hybrid variational solver approach~\citep{peruzzo2014variational,mcclean2016theory,kyriienko2021solving} that can be deployed on NISQ systems. Quantum PINNs are essentially variational quantum circuits using quantum computers for the evaluation of the optimizer's cost function. 

This work focuses on Continuous Variable (CV) quantum computing formulation, an alternative to the popular qubit-based universal quantum computing because CV quantum computing is a more convenient framework for PINN development than qubit-based approaches. CV quantum computing uses physical observables, such as the strength of an electromagnetic field, whose numerical values belong to continuous instead of discrete intervals, like qubit-based quantum computing. In some sense, CV quantum computation is analog in nature, while qubit-based quantum computation is digital. CV quantum computing dates back to 1999, first proposed by~\cite{lloyd1999quantum}. Refs.~\citep{braunstein2005quantum, weedbrook2012gaussian} provide extensive in-depth reviews of CV quantum computing. While the most popular implementations of quantum computers, e.g., IBM, Rigetti, and Google quantum computers, use superconductor transmon qubits, CV quantum computing is implemented with mainly quantum optics~\citep{slussarenko2019photonic} and also ion traps~\citep{ortiz2017continuous}. As the PINN method intends to approximate a continuous function, it is more natural to adopt CV quantum computing than a qubit approach~\citep{knudsen2020solving}. In this work, we extend the work of~\cite{killoran2019continuous}, \cite{knudsen2020solving} and~\cite{kyriienko2021solving} by investigating the performance of quantum PINN exploiting CV quantum neural networks. CV quantum neural networks are better suited than qubit-based quantum computing for performing regression calculations. Because the main PINN usage is for performing regression tasks (instead of classification), CV quantum neural networks are the ideal framework for PINN development. 


The paper is organized as follows. We first briefly review CV quantum computing and neural network and present a design of the quantum PINN together with the experimental setup in Section~\ref{pinndesign}. We discuss the impact of different optimizers, differentiation techniques, quantum neural network depth, and batch size on the PINN performance in Section~\ref{results}. Finally, Section~\ref{conclusion} summarizes the results, discusses the limitations of this work, and outlines future opportunities for quantum PINN development.

\section{Quantum Physics-Informed Neural Networks}
\label{pinndesign}
In this section, we introduce CV quantum computing, its basic gates, and a simple formulation of the quantum neural network unit, the basic building block for CV quantum neural networks. We then discuss the design of a quantum PINN and describe the experimental setup, comprising the programming setup and quantum computer simulator in use.

\subsection{Continuous Variable Quantum Computing and Neural Networks}
\label{cvq}
The CV quantum computing approach is based on the concept of \emph{qumode}, the basic unit carrying information in CV quantum computing. We express the qumode  $\ket{\psi}$, in the basis expansion of quantum states, as 
\begin{equation}
\ket{\psi} = \int \psi(x) \ket{x} dx,
\end{equation}
where the states are the eigenstates of the $\hat{x}$ quadrature, $\hat{x} \ket{x} = x \ket{x}$ with $x$ being a real-valued eigenvalue. This is an alternative formulation to the popular qubit-based approach. In this latter case, the qubit $\ket{\phi} $ is expressed as the combination of the states $\ket{0}$ and $\ket{1}$ as
\begin{equation}
\ket{\phi} = \phi_0 \ket{0} + \phi_1 \ket{1}. 
\end{equation}
While for qubit-based quantum computing, we use a set of discrete coefficients, such as $\phi_0$ and $\phi_1$, in the case of CV based we have a continuous of coefficients (a continuous eigenvalue spectrum), giving the name of this approach. All the quantum computing settings with continuous quantum operators perfectly match CV quantum computing. The position ($ \hat{x}$) and momentum ($ \hat{p}$) operators, constituting the so-called phase space, are good examples of continuous quantum operators we use in this work. The position operator is defined as follows:
\begin{equation}
\hat{x} = \int_{-\infty}^\infty x \ket{x} \bra{x} dx,
\end{equation}
where the vectors $\ket{x}$ are orthogonal. Similarly, the momentum operator is defined as:
\begin{equation}
\hat{p} = \int_{-\infty}^\infty p \ket{p} \bra{p} dp,
\end{equation}
with $\ket{p}$ being orthogonal vectors. A qumode $i$ is associated with a pair of position and momentum operators $(\hat{x}_i,\hat{p}_i)$. These operators do not commute, leading to the Heisenberg uncertainty principle for the simultaneous measurements of $\hat{x}$ and $\hat{p}$.

As in the well-established qubit-based formulation, CV quantum computation can be expressed using low-level gates that can be implemented, for instance, as optical devices. A CV quantum program can be seen as a sequence of gates acting on one or more qumodes. Four basic \emph{Gaussian} gates operating on qumodes are necessary to develop CV quantum neural networks and PINNs. These four gates of linear character are:

\begin{itemize}
\item \textbf{Displacement} Gate - $D(\alpha)$: $\begin{bmatrix} x \\ p \\ \end{bmatrix} \rightarrow  \begin{bmatrix} x + \Re(\alpha) \\ p + \Im(\alpha) \\ \end{bmatrix}$. This operator corresponds to a phase space shift by displacing a complex number $\alpha \in \mathbb{C}$.
\item \textbf{Rotation} Gate -  $R(\phi)$: $\begin{bmatrix} x \\ p \\ \end{bmatrix} \rightarrow \begin{bmatrix} \cos(\phi) & \sin(\phi) \\ -\sin(\phi) & \cos(\phi) \\ \end{bmatrix}   \begin{bmatrix} x \\ p \\ \end{bmatrix} $. This operator corresponds to a rotation of the phase space by an angle $\phi \in [0, 2\pi]$.
\item \textbf{Squeezing} Gate - $S(r)$: $\begin{bmatrix} x \\ p \\ \end{bmatrix} \rightarrow \begin{bmatrix} e^{-r} & 0 \\ 0 & e^r \\ \end{bmatrix}   \begin{bmatrix} x \\ p \\ \end{bmatrix} $. This operation corresponds to a scaling operation in the phase space with a scaling factor $r \in \mathbb{C}$.
\item \textbf{Beam-splitter} Gate - $BS(\theta)$:  $\begin{bmatrix} x_1 \\ x_2 \\  p_1 \\ p_2 \\ \end{bmatrix} \rightarrow  \begin{bmatrix} \cos(\theta) & - \sin(\theta) & 0 & 0  \\ \sin(\theta) & \cos(\theta) & 0 & 0\\  0 & 0 & \cos(\theta) & - \sin(\theta)   \\  0 & 0 & \sin(\theta) & \cos(\theta) \\ \end{bmatrix}    \begin{bmatrix} x_1 \\ x_2 \\  p_1 \\ p_2 \\ \end{bmatrix}   $. This operation is similar to a rotation between two qumodes by an angle $\theta \in [0, 2\pi]$.
\end{itemize}

An important derived gate is the \textbf{interferometer} that can be formulated as a combination of beam-splitter and rotation gates. In the limit of one qumode, the interferometer reduces to a rotation gate. By combining these Gaussian gates, we can define an affine transformation~\citep{killoran2019continuous} that is instrumental for the expression of neural network computation. 

In addition, to these four basic Gaussian gates defined above, non-Gaussian gates, such as the \textbf{cubic} and \textbf{Kerr} gates, provide a non-linearity similar to the non-linearity performed by the activation functions in the classical neural network. Most importantly, the non-Gaussian gates, when added to the Gaussian gates to form a sequence of quantum computing units, provide the universality of the CV quantum circuit: we can guarantee that we can produce any CV state with at most polynomial overhead. In this work, we use the Kerr gate because the CV quantum simulators provide a Kerr gate model that is more accurate than the cubic model. The Kerr gate is often expressed as $K(\kappa)$ and has $\kappa \in \mathbb{R}$ as the quantum gate parameter. 

Finally, an operation's result in quantum computing is a measurement operation. In this work, as a result of the measurements, we evaluate the expected value for the quadrature operator $\hat{x}$: 
\begin{equation}
\bra{\psi_x} \hat{x}  \ket{\psi_x}.
\end{equation}


As mentioned in Section~\ref{intro}, the most promising hardware implementation of CV quantum gates uses photonic technologies. An example of a quantum photonic computer is Xanadu's Borealis quantum computer~\footnote{\url{https://www.xanadu.ai/products/borealis/}}~\citep{madsen2022quantum}, designed for solving Gaussian Boson Sampling (GBS) problems. In this system, a laser source, generated by an Optical Parametric Oscillator (OPO), creates a train of identical light pulses, each constituting a qumode. These qumodes are then injected into a sequence of dynamically programmable loop-based interferometers: the beam splitters and rotation can be programmed to selectively route qumodes into optical delays lines so they can interfere with later qumodes (this technique is called \emph{time-multiplexing}). Finally, the state of the system is measured on the photon number basis, or \emph{Fock state}, employing an array of Photon-Number Resolving (PNR) detectors based on superconducting transition edge sensors. The PNR detectors require cryogenic cooling. The development of CV quantum computing systems is a very active current research area ~\citep{fukui2022building}.

\begin{figure}[h!]
\begin{center}
\includegraphics[width=\textwidth]{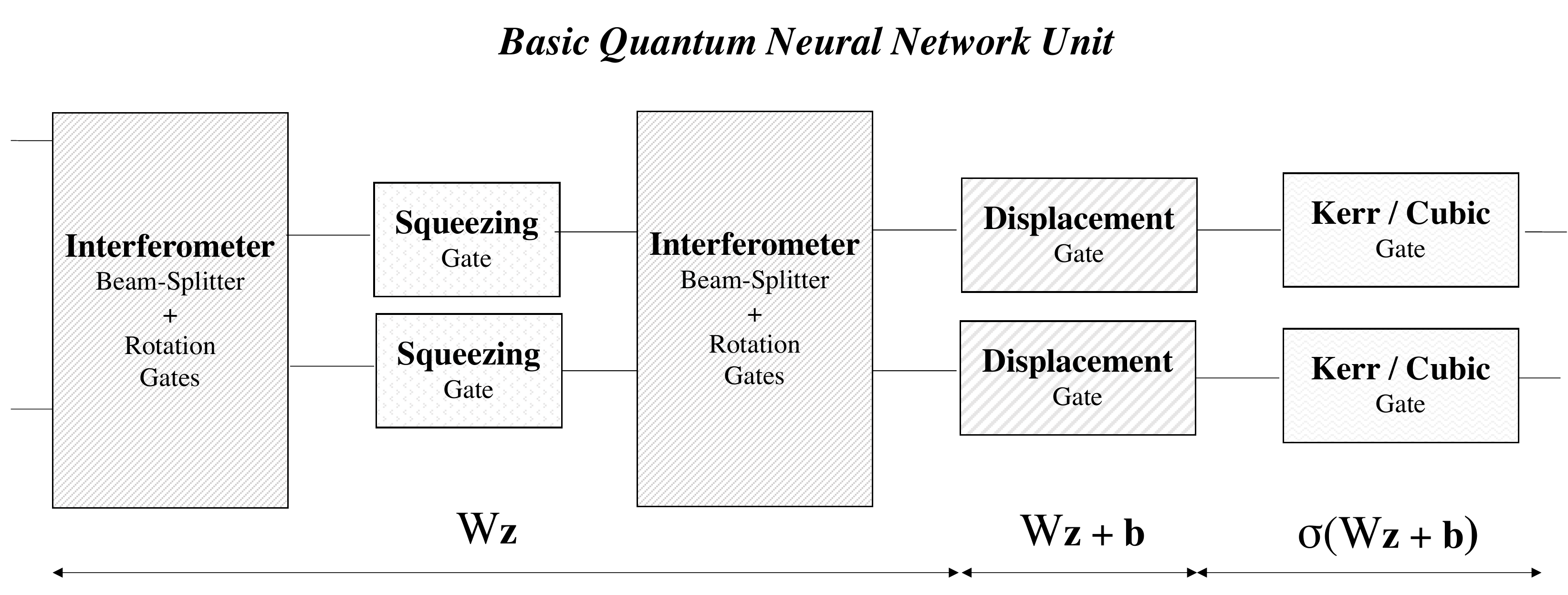}
\end{center}
\caption{A basic fully connected quantum neural network consists of one or more quantum neural units comprising four/five CV quantum gates: interferometer (beam-splitter + rotation), squeezing, displacement, and Kerr or Cubic gates. A quantum neural unit performs a non-linear transformation on an affine operator, similarly to classical quantum neural networks.}\label{fig:QNNunit}
\end{figure}

Equipped with the concepts of CV quantum gates and their parameters and the expected value of the quadrature operator, we can now develop a quantum neural network by following the seminal work by~\cite{killoran2019continuous}. The basic building block of a quantum neural network is the quantum neural network unit (also called a quantum network layer in the literature) which is akin to the classical neural network unit. Figure~\ref{fig:QNNunit} shows the basic components of a quantum neural unit. The first three components of the quantum neural unit are a succession of a first interferometer, a squeezing gate, and a second interferometer. It is shown in Ref.~\citep{killoran2019continuous} that the result of these operations is analogous to multiplying the phase space vector by the neural network weights $W$ (parameters of the interferometers and squeezing gates). As in the classical neural networks, a displacement gate mimics the addition of bias $\mathbf{b}$. Finally, a Kerr gate (or a cubic gate) introduces a non-linearity similar to the activation function $\sigma$ in classical neural networks. 
\begin{equation}
\ket{\mathbf{x}}   \rightarrow \ket{\sigma(W\mathbf{x} + \mathbf{b})}.   
\end{equation}
We can create a quantum neural network by stacking multiple quantum neural units in a sequence. It is important to note that for one qumode, each gate can be controlled by a total of seven gate parameters ($\alpha, \phi, r, \theta$, and $\kappa$) that can be a real-value ($\phi, \theta$, and $\kappa$) or complex number ($\alpha, r$). Two real numbers can express the complex-valued parameters in the Cartesian (with the real and imaginary part) or polar (with amplitude and phase) formulations. The quantum circuit parameters are further divided into \emph{passive} and \emph{active} parameters: the beamsplitter angles and all gate phases are passive parameters, while the displacement, squeezing, and Kerr magnitude are active parameters. The training of quantum neural networks aims at finding the parameter values ($\alpha, \phi, r, \theta$, and $\kappa$) for different qumodes and quantum neural units to minimize the PINN cost function, that is, the PDE residual.

\subsection{Quantum Physics-Informed Neural Networks }
\begin{figure}[h!]
\begin{center}
\includegraphics[width=\textwidth]{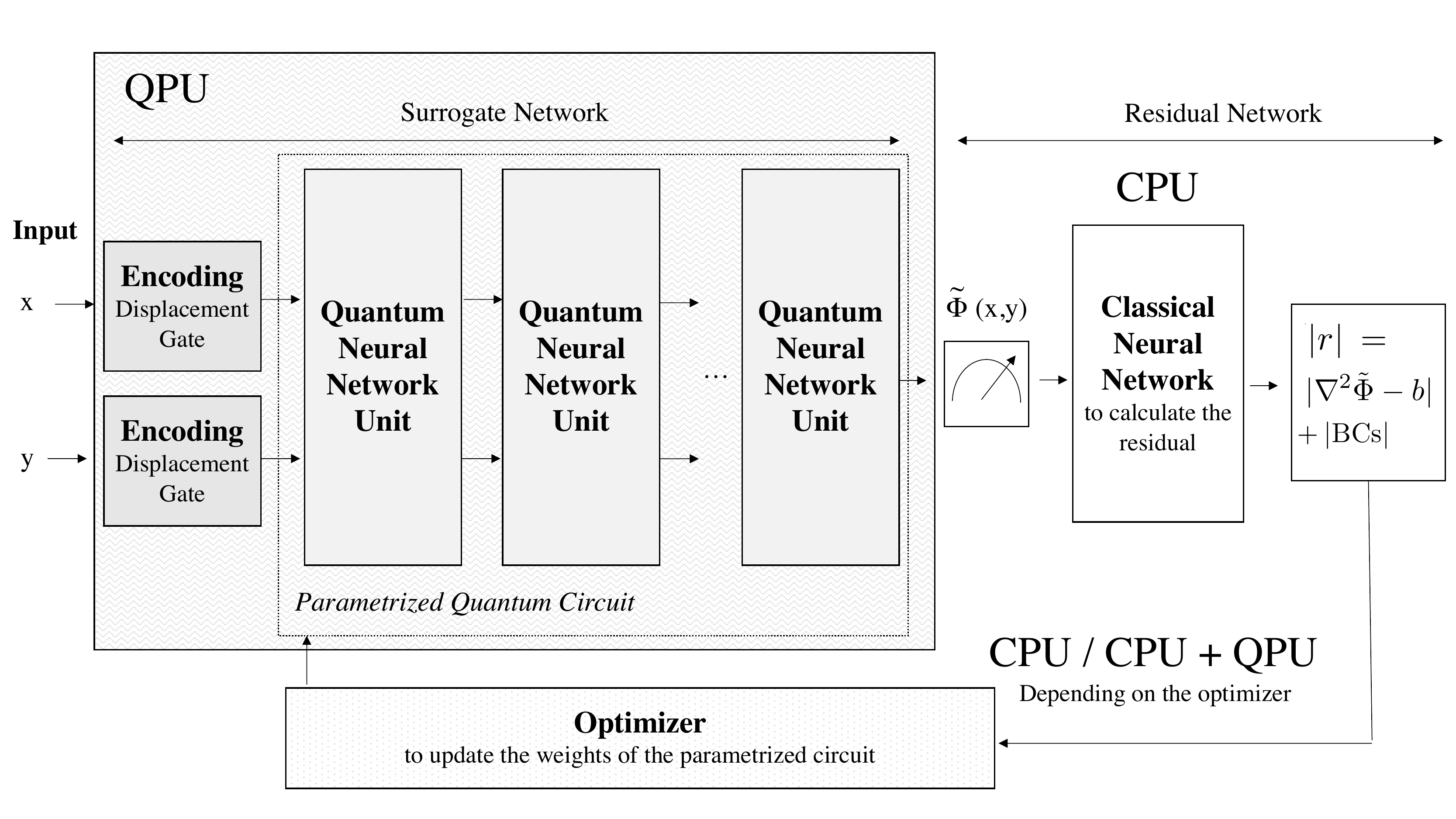}
\end{center}
\caption{A diagram representing the workflow in a quantum PINN. The collocation points values are the input of the quantum PINN and are encoded as a displacement (only with the real part). The PINN surrogate/approximator network is expressed as a parametrized quantum circuit consisting of several connected quantum neural network units. The PINN residual function is calculated on the CPU using automatic differentiation. Some optimizers might require the usage of a quantum computer for the cost function evaluation.}\label{fig:QPINN}
\end{figure}

Quantum PINNs extend the CV quantum neural networks~\citep{knudsen2020solving, kyriienko2021solving}.  Figure~\ref{fig:QPINN} shows an overview of the workflow and resource usage of a quantum PINN for solving a 2D Poisson equation and associated boundary conditions, $\nabla^2 \tilde{\Phi} (x,y) =  b (x,y)$. The Poisson equation is an omnipresent governing equation in scientific computing: electrostatic and gravitational forces are, for instance, governed by the Poisson equation 

As the first step of the quantum PINN, we encode the collocation point as a real-valued displacement of the vacuum state that is the lowest energy Gaussian state with no displacement or squeezing in phase space. One of the significant advantages of using CV quantum neural networks is the ease of encoding the input (the collocation point) into the quantum neural network without requiring normalization. After the initial displacement encodes the collocation point, the qumode feeds into the quantum PINN. A quantum PINN combines two neural networks working together in succession:
\begin{itemize}
\item \textbf{Quantum Surrogate Neural Network.} The quantum neural network surrogate is a CV quantum neural network as described in Ref.~\citep{killoran2019continuous}. The quantum surrogate neural network is a parametrized quantum circuit that takes as input the collocation point coordinates $x$ and $y$, encoded as a displacement of the vacuum state, and gives the approximated solution $\tilde{\Phi}(x,y)$ as the expected value of the quadrature operator. The solution of the PDE is encoded in the parametrized quantum circuit and accessible by running the surrogate quantum surrogate neural network on the QPU. After the training process, we run the quantum circuit giving a displaced vacuum state to calculate the approximated solution.
\item \textbf{Residual Neural Network on CPU.} In a matrix-free fashion, quantum PINNs do not require storing and encoding the problem matrix and the source vector. Instead, the residual network encodes the governing equations (in this case, the 2D Poisson equation). The residual network is not trained, e.g., its weights are not updated, and its only function is to provide the quantum surrogate neural network with a loss function that is the residual function for the domain inner points ($ip$):
\begin{equation}
| r |_ {ip} =|  \nabla^2 \tilde{\Phi} (x,y) -  b (x,y) |. 
\end{equation}
In addition to satisfying the governing equation in the domain inner points, the collocation points on the boundaries must satisfy the problem boundary conditions. For instance, if a problem requires the solution to vanish at the boundaries then a specific residual function for the boundary collocation points ($bp$) is specified as $| r |_ {bp} = | \tilde{\Phi} (x_B,y_B) |$. Traditionally, PINNs use automatic differentiation to calculate the differential operators, such as the nabla operator in the Poisson equation. To perform the automatic differentiation, we rely on the ideal analytical model of the CV quantum gates, as presented in Section~\ref{cvq}, and the chain rule. In addition to automatic differentiation, it is possible to perform numerical differentiation to calculate the differential operators. However, this requires introducing a computational grid (with the solutions calculated only on discrete points) and a discretization of the differential operators. The calculation of the residual function occurs only on the CPU and does not involve the QPU.
\end{itemize}

In the quantum PINN, the approximated solution of the quantum neural network is used to calculate the residual function. The absolute value of the residual function, including the boundary conditions, constitutes the loss function $\mathcal{L}$. During this work, we found that performing the operations in batches, e.g., running in parallel with the quantum PINN with different collocation points and averaging the loss function over the batch, improves the quality of the results in terms of smoothness of the solution and the computation performance of the quantum simulator. For this reason, as cost function, we take the average of cost functions evaluated at the different batch collocation points:
\begin{equation}
\mathcal{L} = \frac{ \sum_i^{B_s}  | r  |_ {ip} } { B_s}  + \frac{ \sum_i^{B_s}  | r |_ {bp}   } { B_s } = \frac{ \sum_i^{B_s}  |  \nabla^2 \tilde{\Phi} (x_i,y_i) -  b (x_i,y_i) |} { B_s}  + \frac{ \sum_i^{B_s}  |\tilde{\Phi} (x_{i,B},y_{i,B}) | } { B_s },
\end{equation}
where $B_s$ is the batch size, $(x_i,y_i)$ are random inner collocation points and $(x_{i,B},y_{i,B})$ are boundary collocation points. To run the quantum neural in a batch is equivalent to running it using multiple qumodes. For instance, if the quantum neural network uses only one qumode, e.g., a one-dimensional Poisson equation problem, with a batch size of 32, we can run the full batch using 32 qumodes on the QPU.

The updates of the quantum surrogate neural network parameters are determined by running a stochastic optimizer that relies on the calculation of gradient (and Hessian for second-order optimizers) and often on a learning rate, e.g., a step size toward a minimum of a loss function. Stochastic optimizers are \emph{adaptive} if the learning rate change during the optimizer iterations. Examples of adaptive optimizers are RMSprop, Adam, Adam with Nesterov momentum (Nadam)~\citep{ruder2016overview}, and Adadelta~\citep{zeiler2012adadelta}. If the calculation of the gradient uses automatic differentiation, then the optimizer only runs on the CPU; in the case of numerical differentiation, such as in the case of Simultaneous Perturbation Stochastic Approximation (SPSA) and L-BFGS-B, the optimizers use both CPU and QPU, called for the function evaluation with the quantum surrogate neural network.

\subsection{Quantum Neural Network Implementation}
For the design and development of the Quantum PINN, we use Xanadu's \texttt{Strawberry Fields}~\footnote{\url{https://strawberryfields.ai/}} CV programming and simulation framework \cite{killoran2019strawberry, bromley2020applications}. Different backends to simulate the CV quantum computers are available in Strawberry Field, including the \texttt{fock}, \texttt{tf}, and \texttt{gaussian} backends. The \texttt{fock} and \texttt{tf} backends express the modes' quantum state with the Fock or particle basis. Arbitrary quantum states are expressed up to a photon cutoff when using these backends. However, this comes with an exponential increase in space complexity as a function of the number of modes with the base of the exponent scaling. This constrains the quantum simulator memory usage that critically depends on the number of qumodes and the cutoff number. The \texttt{gaussian} backend does not suffer the problem of exponential scaling in the memory required. However, only a subset of states can be represented.

In this work, we use the TensorFlow backend  \cite{killoran2019strawberry} that provides the quantum computer simulator and an expressive domain-specific language to formulate CV quantum neural network calculations. The TensorFlow backend represents the quantum state of the modes using the Fock or particle basis, and it is subject to the memory constraints of representing quantum states with the Fock or particle basis. The advantage of using the TensorFlow backend is that the programmer can use all existing Keras and TensorFlow automatic differentiation, optimizers, and tensor operation.

While both the \texttt{Strawberry Fields} and \texttt{TensorFlow} interfaces will be subject to changes in the future, we show some snippet code to demonstrate the simplicity of developing a quantum PINN with these programming interfaces. The \texttt{Strawberry Fields} quantum computer simulator is easily initialized with:
\begin{lstlisting}[language=Python,numbers=none]
eng = sf.Engine(backend="tf", backend_options={"cutoff_dim": 125, "batch_size": n_collocation_points})
prog = sf.Program(1)
\end{lstlisting}
Note that we select the \texttt{tf} backend, a cutoff for representing state equal to 125 and batch size equal to the number of collocation points we use in the PINN. Our code only uses one qumode set with \texttt{sf.Program(1)}.

To express our CV quantum circuit, we use \texttt{Blackbird}, which is the Assembly language built into \texttt{Strawberry Fields}. The \texttt{Strawberry Fields} framework allows us to express the quantum circuit, including the set of gates for the implementation of the quantum neural network, in a straightforward form by providing a sequence of gates acting on a qumode $q$. The quantum circuit for encoding the collocation point and executing a forward pass with a one-layer network using one qumode (in this case, the beam-splitter reduces to a rotation) is shown in the Listing of Python code below. 

\begin{lstlisting}[language=Python,numbers=none]
#    QPINN Surrogate Circuit: Input + One Quantum Unit / Layer               
with prog.context as q:
    #  Initially the qumode q is in the vacuum state 
    Dgate(x) | q    #  Encode the collocation point x with a displacement on q
    # One quantum neural unit to express the surrogate network
    Rgate(r1) | q  # Beam-splitter 1: beam-splitter reduces to rotation with one qumode 
    Sgate(sq_r1, sq_phi1) | q # Squeezeer 
    Rgate(r2) | q # Beam-splitter 2: beam-splitter reduces to rotation with one qumode
    Dgate(alpha1, phi1) | q   # Displacement (similar to adding bias in classical neural network)
    Kgate(kappa1) | q # Kerr gate: non-linear transformation (similar to activation function in NN)
\end{lstlisting}

Note that in a simple setup of one quantum neural unit and one qumode we have seven neural network weights (circuit parameters): \texttt{r1} for the first rotation, \texttt{sqr\textunderscore r1} and \texttt{sqr\textunderscore phi1} for the amplitude and phase of the squeezer,  \texttt{alpha1} and \texttt{sqr1} for the real and the imaginary parts  for the displacement and \texttt{k1} for the Kerr gate. In the case of a network with one quantum neural unit and one qumode, the seven network weights (quantum circuit parameters) are optimized to minimize the PINN residual function. With one qumode, a network with one, two, three, and four quantum neural units will require to optimize seven, 14, 21, and 28 weights, respectively.

The snippet of Python code below shows the implementation of executing the quantum circuit defined above, using the \texttt{run()} method, extracting the resulting state, and obtaining the expected value of the quadrature operator with  \texttt{quad\_expectation()} method. To calculate the second-order derivative of the Poisson equation for the residual function, we first identify the region of code defining the operations that are differentiated using the TensorFlow 2 \emph{gradient taping} and then perform the differentiation with the \texttt{gradient()} method.
\begin{lstlisting}[language=Python,numbers=none]
def QPINNmodel(...):   
    if eng.run_progs:
       eng.reset()
    x_in = tf.Variable(np.random.uniform(low=0.0, high=Lx, size=n_collocation_points)) # Random points
    with tf.GradientTape() as tape2:
         with tf.GradientTape() as tape1:
              result = eng.run(prog, args={"x": x_in, "alpha1": alpha1_in, "phi1": phi1_in, 
              "r1": r1_in, "sq_r1": sq_r1_in, "sq_phi1": sq_phi1_in, "r2": r2_in, "kappa1": kappa1_in})
              state = result.state
              mean, var = state.quad_expectation(0)
     # calculate the second order derivative with automatic differentiation
         dudx = tape1.gradient(mean, x_in)
     du2dx2 = tape2.gradient(dudx, x_in)
     b =  ...  # define the known term  
     res = du2dx2 - b      # define the residual
     lossIP = tf.reduce_mean(tf.abs(res)) # take the mean over the batch (IP = Inner points)
     # now calculate the residual for the collocation points on the boundaries
     ...
     loss = lossIP + lossBC1 + lossBC2
     return loss
...
# calculate one optimization step
with tf.GradientTape() as tape:
  loss = QPINNmodel(...)  
opt = tf.keras.optimizers.SGD(learning_rate=rate, name="SGD")  # pick the optimizer
gradients = tape.gradient(loss, [alpha1_in, phi1_in,r1_in, sq_r1_in, sq_phi1_in,r2_in, kappa1_in])
opt.apply_gradients(zip(gradients, [alpha1_in,phi1_in,r1_in, sq_r1_in, sq_phi1_in, r2_in,kappa1_in ]))
...
\end{lstlisting}
We can calculate the loss function using the result from the differentiation and the value of the known term at the collocation point. Note that the operation is performed in parallel on all the collocation points, so we take an average of the absolute value of the residual with \texttt{tf.reduce\_mean(tf.abs())}. To update the quantum network parameters, we calculate the first-order derivative of the loss function with respect to the quantum circuit parameters and use these values to update them with \texttt{apply\_gradients()} operation. In this particular case, we use the \texttt{SGD} Keras optimizer.

\subsection{Experimental Setup and Accuracy Metrics}
In this study, we rely on the Python Strawberry Fields CV quantum simulator and a series of Python modules to enable efficient vector calculations (on the CPU) and additional optimizers not included in TensorFlow / Keras framework. We use Python~3.10.4, NumPy (1.22.4), and SciPy (1.8.1) modules. We perform the experiments using the quantum computer simulator provided by Strawberryfields framework, version~0.22.0. In all the simulations, we use one quantum mode and a cutoff dimension of 125 for the Fock basis. We check each measurement's state vector's norm to verify that quantum computer simulation is accurate. Neural networks and quantum computer simulators inherently comprise a level of stochasticity. For this, we set random number generator seeds for TensorFlow and NumPy.

As part of this study, we evaluate the usage of several optimizers: RMSprop, Adam, Adam with Nesterov momentum (Nadam)~\citep{ruder2016overview}, Adadelta~\citep{zeiler2012adadelta}, Simultaneous Perturbation Stochastic Approximation (SPSA)~\cite{spall1998overview}, and the Limited-memory Broyden–Fletcher–Goldfarb–Shanno (L-BFGS-B)~\citep{liu1989limited} optimizers. We use the TensorFlow~2.9.0  SGD, RMSprop, Adam, Nadam, and Adadelta implementations and automatic differentiation capabilities.  For the SPSA optimizer, we use its implementation, available at \url{https://github.com/SimpleArt/spsa} that provides an adaptive learning rate. Finally, we use SciPy~1.8.1 L-BFGS-B optimizer in combination with the TensorFlow SGD.

For the sake of simplicity, to evaluate the quantum PINN, we use a simple one-dimensional Poisson's problem with Dirichlet boundary conditions fixed to zero: $d^2 \Phi / d x^2 = b(x)$. For testing purposes, we choose kinds of sources (the $b$ term in the Poisson equation) and two domain sizes:
\begin{enumerate}
\item \textbf{Quadratic}:  $b(x) = x (x - 1)$,    [ 0, 1 ],   $ \Phi(0) = 0,  \Phi(1) = 0$. The solution, in this case, is a parabola with the first derivative equal to zero at the center of the domain, $x = 0.5$.
\item \textbf{Sinusoidal}:  $b(x) = \sin(2 x)$,  [ 0,  $2\pi$],   $ \Phi(0) = 0,  \Phi(2\pi) = 0 $. This is a more challenging test case as the solution has four points where the first derivative zeros.
\end{enumerate}
An extension of quantum PINN to solve a two-dimensional would require to use of two qumodes to initially encode the $x$ and $y$ coordinates of the collocation points and having interferometers (instead of the simple rotation gate in the case of one qumode) in the quantum neural network to entangle the two qumodes. In the two-dimensional, the residual network encodes the Laplacian operator instead of the one-dimensional derivative.

The \emph{baseline} quantum neural network hyper-parameters are the following: we set the learning rate equal to 0.01 and 0.0001 for the quadratic and sinusoidal source terms cases, respectively. Optimizers' performance highly depends on the initial learning rate. We completed a grid search for setting the learning rate for SGD as it used a fixed learning rate during the training and was more susceptible to the exploding and vanishing gradient problem. We perform 500 optimizer iterations. The collocation points are drawn randomly within the PDE domain at each iteration. The number of collocation points per is equal to the batch size. As baseline runs, we choose 32 for the batch size. For the boundary conditions, we also evaluate the cost functions at the boundaries with a number of collocation points equal to batch size, e.g., 32 for the baseline cases. The quantum circuit parameters are initialized with a normal distribution centered at zero and a standard deviation of 0.05. The simulation of a QPINN implemented with Strawberry Fields takes approximately twenty minutes on a modern laptop.

To characterize the accuracy of the quantum PINN, we check the loss function values (the absolute value of the residual) and the final error norm after the training. To evaluate the final error, we compare the analytical solution and quantum PINN results using a uniform grid of 32 points (including the boundary points) and take the Euclidean norm of the PINN approximated solution minus the analytical solution on the 32 points.

\section{Results}
\label{results}
\begin{figure}[h!]
\begin{center}
\includegraphics[width=\textwidth]{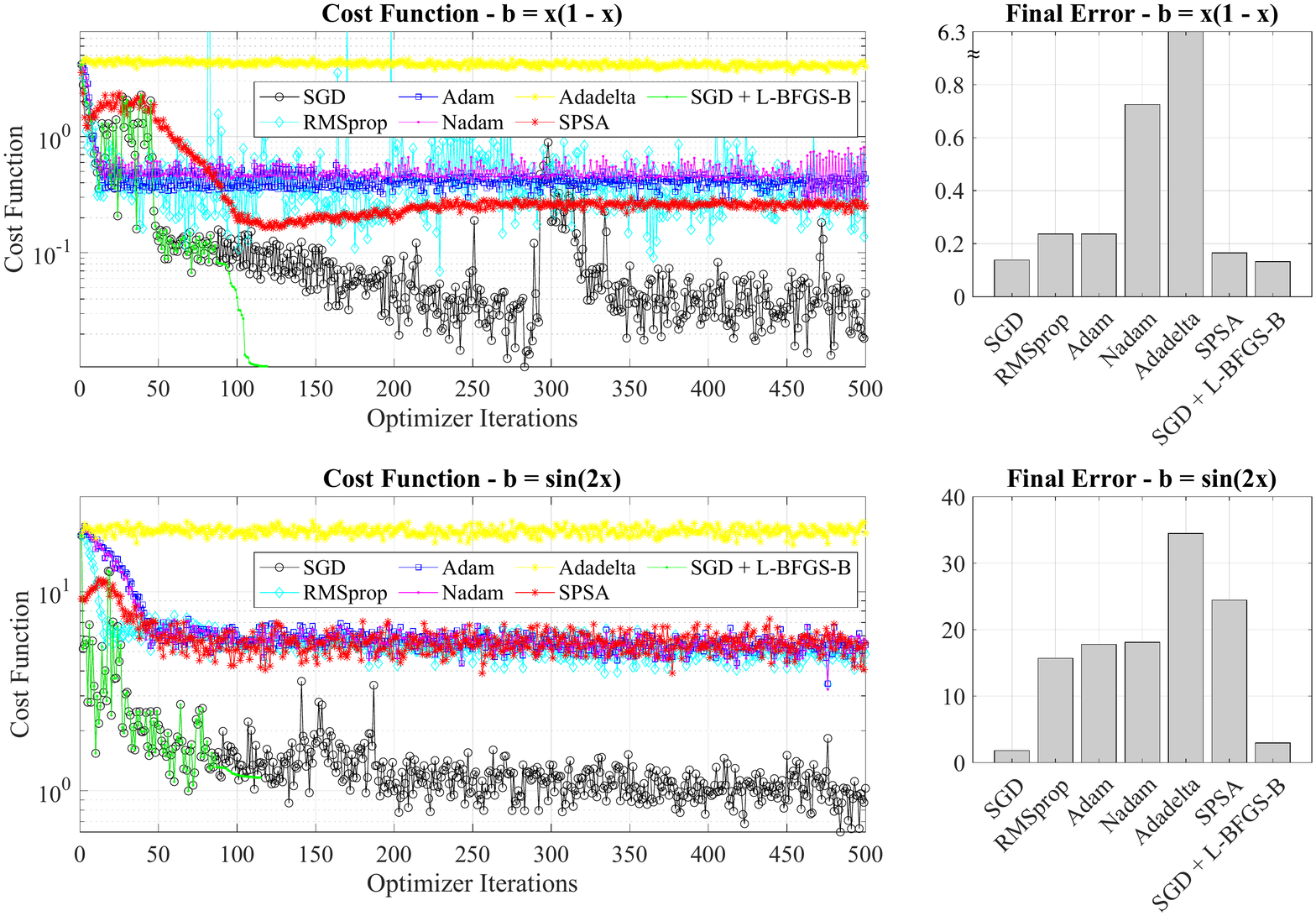}
\end{center}
\caption{Cost function value evolution and final errors for different stochastic optimizers (SGD, RMSprop, Adam, Nadam, Adadelta, SPSA, and SGD + L-BFGS-B). The SGD optimizer has the best performance for both test problems.}\label{fig:optimizers}
\end{figure}

As the first step in our experiments, we investigate the accuracy of different stochastic optimizers for the two one-dimensional Poisson problems with the quadratic and sinusoidal sources. Figure~\ref{fig:optimizers} shows the cost function value for the two test problems on the left panels, while the right panels show the final error after the training. 

By analyzing the cost function value and final error, it is clear that the SGD optimizer outperforms the adaptive (RMSprop, Adam, Nadam, and Adadelta) and SPSA optimizers. In general, we find that adaptive optimizers, such as Adam, tend to converge to local minima in the training landscape without exiting. In the case of the Adadelta optimizer, we do not observe convergence to the solution.  On the other hand, a noisier optimizer, such as SGD, can escape the local minima and better explore the optimization landscape. For instance, by analyzing the cost function value for the SGD optimizer in the parabolic source case (black line in the top left panel of Figure~\ref{fig:optimizers}), we note that the optimizer escapes a local minimum approximately after 300 iterations. The SPSA optimizer also allows us to provide additional noise to hop between different convergence basins, possibly achieving a similar behavior of SGD. However, we find that SPSA does not perform better than SGD. While all the optimizers perform relatively well with the parabolic case (they all capture the parabolic nature of the solution), the adaptive optimizers fail to recover the sinusoidal nature of the solution in the second test case (not shown here) and leading to significant final errors (see the bottom right panel in Figure~\ref{fig:optimizers}). 

In classical PINN, a second-order optimizer, L-BFGS-B, is used after the Adam optimizer to speed up the PINN convergence, thus requiring considerably fewer iterations~\cite{raissi2019physics, markidis2021old}. In classical PINN, L-BFGS-B is not used from the start of the training, as it would quickly converge to a local minimum of the training landscape without escaping it. As in the classical case, we deploy an L-BFGS-B optimizer after 80 optimizer iterations. While L-BFGS-B can reduce the cost function evaluation in both cases, we note that the final error is approximately the same for plain SGD optimizers and SGD combined with L-BFGS-B. As additional tests, we also implemented a multi-step SGD and L-BFGS-B, e.g., a succession of SGD and L-BFGS-B, without achieving a final performance improvement. In classical fully-connected PINNs, Adam and L-BFGS-B optimizers are widely employed and successful for PINNs~\cite{raissi2019physics, markidis2021old}, while in quantum PINN SGD provides better performance than Adam and L-BFGS-B. In quantum PINN, the optimization landscape is more diverse, and its optimizer exploration is more challenging.

One of the main advantages of using PINNs is to leverage Automatic Differentiation (AD) for approximating with high accuracy a derivative at an arbitrary point in the domain. This is a powerful and flexible mechanism for evaluating the residual function in PINN. However, we note that it is possible to calculate the residual function using a fixed number of collocation points, e.g., the nodes of a uniform grid, and approximate the derivative using a Finite Difference (FD) approximation. A similar approach is used in classical fractional PINN~\cite{pang2019fpinns}. This method comes with the disadvantage of using a fixed grid point, hurting the generalization of the solution to different collocation points and expressing the calculations on finite difference stencils. 
\begin{figure}[h!]
\begin{center}
\includegraphics[width=\textwidth]{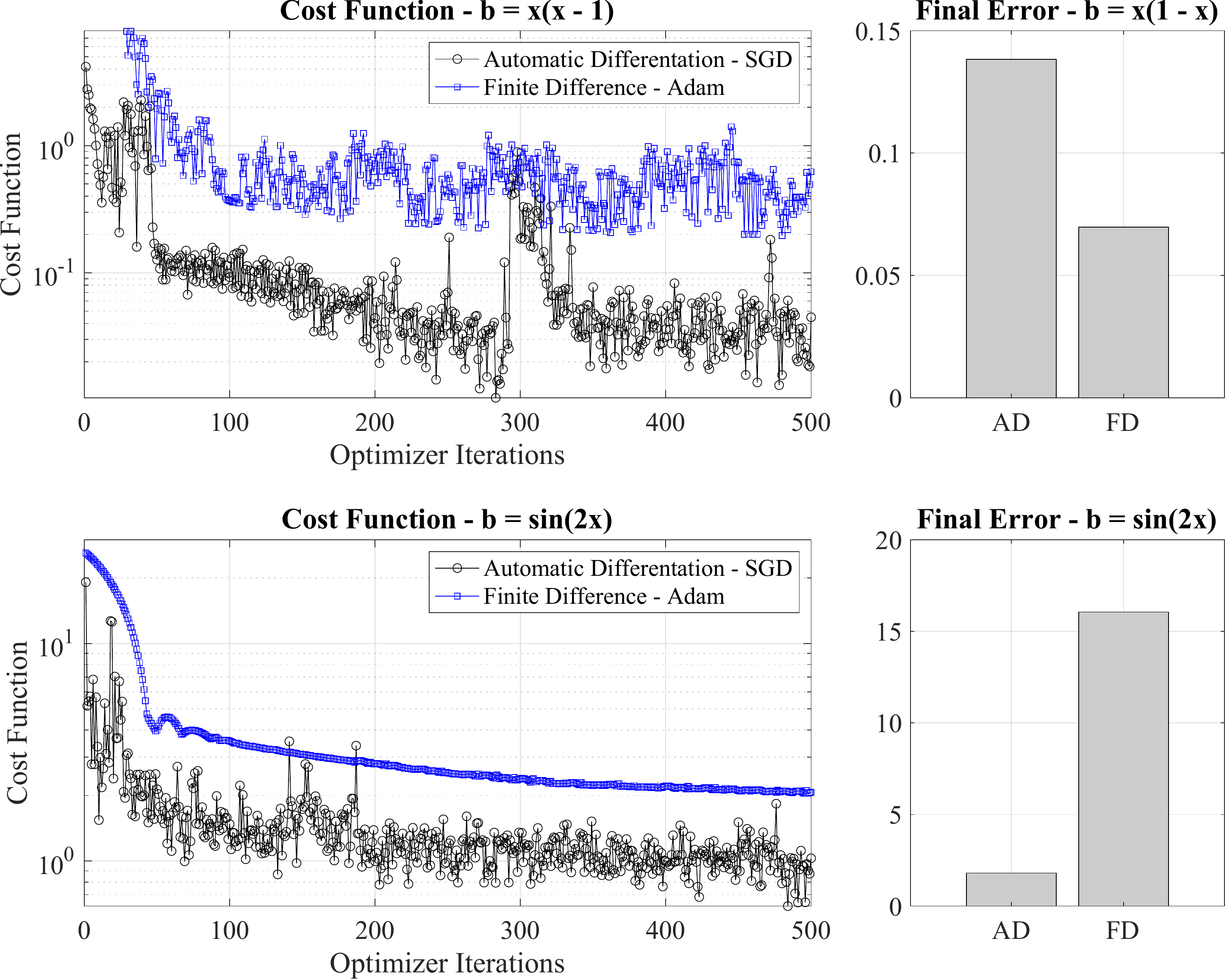}
\end{center}
\caption{Comparison of function cost value (left panels) and final error (right panel) between quantum neural network using Automatic Differentiation (AD) and a Finite Difference (FD) approximation for the calculation of derivatives in the PDE. In the FD case, we use an Adam optimizer (the SGD solver shows high variability in the loss function) and constant uniform collocation point distribution.}\label{fig:ADFD}
\end{figure}

We compare the cost function value evolution and final error for PINN residual function, calculated with AD and FD in Figure~\ref{fig:ADFD}. When using the SGD optimizer and an FD formulation for the residual function, we observed high variability of the cost function values and the occurrence of two-three characteristic cost function values, signaling that the quantum PINN is hopping quickly between a small number of local minima. Using an Adam optimizer, we have a smoother cost function value evolution behavior. For this reason, for the FD case, we switch to an Adam optimizer. From analyzing Figure~\ref{fig:ADFD}, we see that AD provides a lower final error than the quantum PINN using AD for the residual function in the case of the quadratic source case, despite the higher loss function values. However, FD fails (significant final error) in capturing the sinusoidal nature of the solution in the second test case. Because of the use of fixed collocation points, the quantum PINN with FD cannot escape a local minimum leading to a large final error. 

\begin{figure}[h!]
\begin{center}
\includegraphics[width=\textwidth]{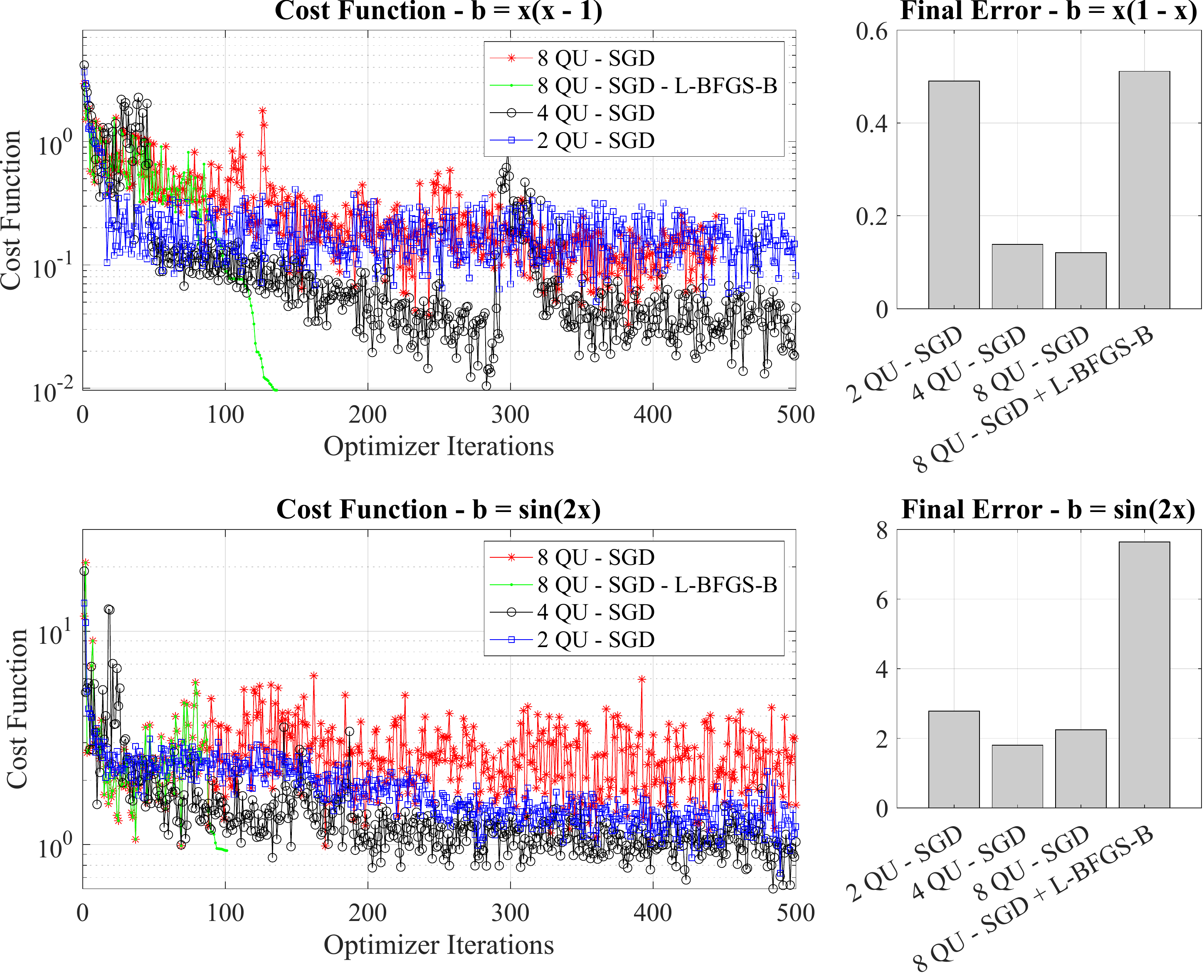}
\end{center}
\caption{Cost function value evolution and final errors for different quantum neural network depths (two, four, and eight quantum neural units).}\label{fig:depth}
\end{figure}

As an additional investigation, we study the impact of the quantum neural network depth (also the depth of the quantum circuit) on the calculations by varying the number of quantum neural network units in the surrogate network. The results of these experiments in terms of the cost function values and final error are presented in Figure~\ref{fig:depth}.

Overall, we find that quantum PINNs with four and eight quantum units have a comparable final error. In general, the network with eight quantum units exhibits a higher variability in the cost function evolution (see the red lines on the left panels of Figure~\ref{fig:depth}). The neural network with two quantum units is too shallow to express the solution at higher accuracy. The interesting point is that in deep PINNs, e.g., with eight quantum neural units, the L-BFGS-B leads to significant error as it converges to local minima of the training landscape with an incorrect solution. This is probably due to switching too early to L-BFGS-B (after 85 SGD iterations) without letting the optimizers increase the exploration of the landscape. 

\begin{figure}[h!]
\begin{center}
\includegraphics[width=\textwidth]{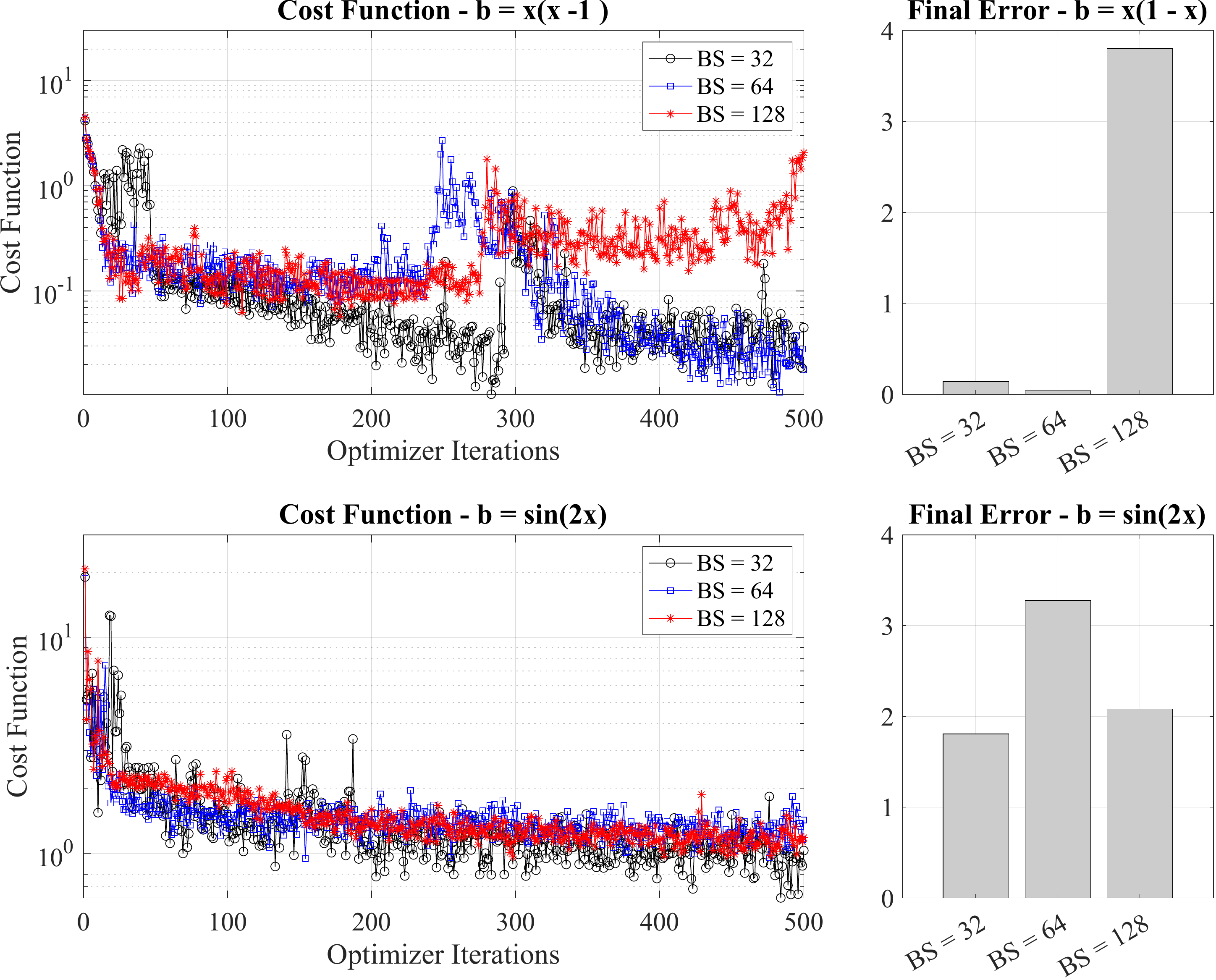}
\end{center}
\caption{Cost function value evolution and final errors for different batch size (32, 64 and 168)}\label{fig:bs}
\end{figure}

Finally, we investigate the impact of the batch size per iteration and show it in Figure~\ref{fig:bs}. The increase in batch size leads to a rise in the number of collocation points and associated cost function evaluations per optimizer iterations.

While intuitively, we would expect a noticeable improvement of the quantum PINN network when increasing the batch size, by analyzing Figure~\ref{fig:bs}, in practice, we do not note a significant impact of the batch size on the performance of the quantum PINN. This is except for the quadratic source case and batch size equal to 128 (red line in the top left panel of Figure \ref{fig:bs}, where the performance drops due to convergence to a local minimum of the training landscape.

\section{Discussion and Conclusion}
\label{conclusion}
This work investigated the development of PINN for solving differential equations on heterogeneous CPU-QPU systems using a CV quantum computing approach. Quantum PINN solvers are fundamentally variational quantum circuit solvers~\citep{cerezo2021variational} with an additional classical residual network to provide a cost function for the optimization step. The evaluation of the approximated solution is always carried out on the quantum computer (in this work, a simulated one). Currently, the calculation of the loss function via the residual neural network is carried out on the CPU. While this study uses a residual network on the CPU, the automatic differentiation might likely be implemented on quantum hardware. Given the gate formulation in Section~\ref{cvq}, differential operators might be automatically synthesized as additional gates to express differentiation.

Overall, we found the programming interfaces conveniently abstract the QPU and quantum hardware: the programmer does not explicitly take care of the data movement or offloading of the quantum circuit execution. As in the case of GPU, the Strawberry Fields and TensorFlow frameworks provide a convenient approach for the transparent usage of the QPU co-processors. In particular, the option of having a batch operation abstracts the parallel operation of multiple parallel cost function evaluations and measurements similarly to GPU. The programmer needs a basic understanding of expressing quantum circuits using Quantum Assembly languages, such as Blackbird. 
 
We showed that a CV quantum neural network could provide a surrogate neural network for approximating a PDE solution to a certain degree in two test cases, solving the one-dimensional Poisson equation with quadratic and sinusoidal source terms. Our results showed that the optimizer's choice is the most impactful on the PINN solver accuracy: SGD solvers lead to a more accurate and stable solution than adaptive optimizers. The depth of the quantum neural network affects the PINN performance. In the two test cases, we found four layers provided higher accuracy than a two-layer quantum network and comparable performance to the eight-layer network. In the two test cases, we did not find a dependency on batch size, e.g., the number of collocation points we need for the training.

The main priority for developing further quantum PINN is to address their current low accuracy results. In all our tests, in practice, reducing the error below a certain value or increasing the convergence in a finite number of iterations has been challenging. This is likely related to the \emph{barren plateau} problem~\citep{mcclean2018barren}, affecting all the hybrid quantum-classical algorithms involving an optimization step: the classical optimizer is stuck on a barren plateau of the training landscape with an exponentially small probability of exiting it. The \emph{barren plateau} problem is fundamentally due to the geometry of parametrized quantum circuits (our quantum surrogate neural network) and training landscapes related to hybrid classical-quantum algorithms~\citep{mcclean2018barren, arrasmith2021effect}. When comparing quantum to classical PINN in our implementation, we found that the optimizer exploration of the training landscape in the case of quantum PINN is not as effective as in classical PINNs, and adaptive and high-order optimizers are less performant than basic SGD optimizers. Potential \emph{classical} strategies to mitigate this problem are the usage of a quantum ResNet or convolutional surrogate quantum networks~\citep{killoran2019continuous, mcclean2018barren}, skip connections~\citep{li2018visualizing}, dropout techniques (in multi-qumodes neural networks)~\citep{srivastava2014dropout}, a structured initial guess, as used in quantum simulations, and pre-training segment by segment~\citep{bengio2006greedy}. 

This exploratory work has several limitations. First, this work does not prove any quantum advantage but shows challenges and opportunities for implementing PINNs on quantum computing systems. The current problem (one-dimensional Poisson equation) does not exploit quantum entanglement. A PINN for solving higher-dimensional problems or systems of equations would require the usage of entangled qumodes. Second, for CV quantum computing, the QPINN has only been tested on a quantum computer simulator, the Strawberry Fields TensorFlow backend. To execute on future photonic quantum computers, additional work must be done to consider the connectivity of the qumodes and hardware constraints, e.g., availability and the performance of gates on the given quantum systems~\citep{watabe2021quantum}. Despite all these limitations, this work lays down a starting point for developing PINNs for quantum computers showing the difference between classical and quantum PINNs and pinpointing current challenges to be addressed for PINN solvers on hybrid QPU-CPU systems.

\section*{Acknowledgments}
The author would like to thank Felix Liu and Pawel Herman for the insightful suggestions about stochastic optimizers and Vincent Elfving for discussing previous fundamental work on DQCs and quantum PINNs.


\bibliographystyle{acm} 
\bibliography{QPINN_arxiv}


\end{document}